\documentclass[pra, twocolumn,
 superscriptaddress,
 amsmath,amssymb,
 aps,
floatfix, 
longbibliography
]{revtex4-1}
\usepackage{graphicx}
\usepackage{bm}

\usepackage{xcolor}
\usepackage[colorlinks=true,citecolor=blue,linkcolor=magenta]{hyperref}
\usepackage{orcidlink}

\usepackage[latin9]{inputenc}
\usepackage{amsmath}
\usepackage{amssymb}
\usepackage{graphicx}
\usepackage{verbatim}
\usepackage{bm}
\usepackage{color}

\usepackage{graphicx,epsfig,amsfonts,amssymb}
\usepackage{bm}
\usepackage{times}
\usepackage{lipsum}
\usepackage{verbatim}

\newcommand{\be}{\begin{eqnarray}}
\newcommand{\ee}{\end{eqnarray}}

\newcommand{\bs}{\begin{equation}\begin{split}}
\newcommand{\es}{\end{split}\end{equation}}

\date{\today}

\begin{document}

\title{Geometric phase in anisotropic Kepler problem:  Perspective for realization in Rydberg atoms}

\author{Nikolai A. Sinitsyn\orcidlink{0000-0002-0746-0400}}\email{nsinitsyn@lanl.gov}
\affiliation{Theoretical Division, Los Alamos National Laboratory, Los Alamos, New Mexico 87545, USA}

\author{Fumika Suzuki\orcidlink{0000-0003-4982-5970}}\email{fsuzuki@lanl.gov}
\affiliation{Theoretical Division, Los Alamos National Laboratory, Los Alamos, New Mexico 87545, USA}
\affiliation{Center for Nonlinear Studies, Los Alamos National Laboratory, Los Alamos, New Mexico 87545, USA}

\begin{abstract}

    We predict a gyroscopic effect that  can be demonstrated with Rydberg atoms   following the dynamics of a Kepler Hamiltonian with an additional uniaxial anisotropy induced by optical ponderomotive force. This effect is analogous to the rotation of the Foucault pendulum in response to the Earth's rotation. We argue that in Rydberg states with a large principal quantum number a similar geometric angle can be generated by mechanical rotations of an atomic-optical setup on time scales between $1~\mu$s and $1~$ms.
\end{abstract}
\maketitle

{\it Introduction}.
Rydberg states of alkali atoms can exhibit dynamics described by the classical Kepler problem without visible quantum corrections \cite{Yeazell1989}. An elliptical electron orbit with principal quantum number $n = 100$ has an orbital size of approximately $d \sim 0.5\ \mu$m, resulting in a large electric dipole moment. This explains the high sensitivity of Rydberg states to external fields \cite{Dutta2007}.  Rydberg atoms are already used as radio-frequency (RF) field sensors operating in the MHz-THz frequency range \cite{artusioglimpse2022modern}.
They are also candidates for future gyroscopic sensing. 

Gyroscopes are devices used for measuring rotation, such as in airplanes and rockets. Traditionally used optical and mechanical gyroscopes are  limited by their response times on the order of a millisecond (ms) or larger \cite{Yazdi2017Gyroscope}, while Rydberg atoms may detect changes in external fields with a temporal resolution potentially faster than a nanosecond (ns). They could unlock new regimes of inertial sensing on short time scales. Although this  has not yet led to a commercial technology, the proof-of-principle experiments have been performed, demonstrating rotational sensing using quantum interference effects in ultracold atoms \cite{anderson2016rydbergsensors}. Similar research has also sparked interest in highly sensitive gravitational sensors based on matter-wave interferometry and has contributed to advancements in fundamental tests of physics \cite{Gustavson1997,krzyzanowska2023matter,durfee2006long,fein2020coriolis,di2021gravitational}.

In this Letter, we propose a different approach to using Rydberg states for measuring mechanical rotation.  We take advantage of  {\it super-integrability} of the Kepler problem, which enables the geometric phase effects, resembling  the rotation of the Foucault pendulum but now on sub-millisecond time scales. These effects modify the electronic Kepler orbits in response to mechanical rotation of the entire setup. At static conditions, after the rotation, such changes are preserved in the Rydberg states during the lifetime $\sim1$~ms \cite{PhysRevX.14.021024,PhysRevResearch.2.022032}.

{\it Origin of geometric phase}. We recall that the ability of the Foucault pendulum to sense rotational motion relies on two key features \cite{Suzuki2025Geometric}. The first is the degeneracy of oscillation frequencies along the $x$ and $y$ axes for pendulum motion confined to the $xy$-plane, which is tangential to the Earth's surface. This property renders the Foucault pendulum a super-integrable system, in which the action variables
\begin{equation}
I_{x} = \frac{1}{2\pi}\oint p_{x} \,dx, \quad I_{y}=\frac{1}{2\pi} \oint p_y\, dy
\label{actions1}
\end{equation}
may vary individually even under adiabatic evolution of system parameters.  The only true adiabatic invariant in such an oscillator is the sum $I=I_x+I_y$. Slow  parameter changes  may still lead to relative changes in $I_x$ and $I_y$, which manifest as a rotation of the oscillation direction within the $xy$-plane. According to  Landau and Lifshitz's volume {\it Mechanics}  \cite{landau1976mechanics}, this behavior is not typical. For incommensurate oscillation frequencies, each of the actions in Eq.~(\ref{actions1}) would instead be an independent adiabatic invariant.

The Kepler problem for an electron of a Rydberg atom, placed at ${\bf R}=0$, is defined by a classical Hamiltonian, 
$
H={\bf p}^2/2m -Q/|{\bf r}|
$, 
where ${\bf r}$ and ${\bf p}$ are the electron's position and momentum vectors in 3D-space, $m$ is electron mass, and $Q$ characterizes the strength of the attractive Coulomb potential to the nucleus. This Hamiltonian also has the super-integrability property: Electronic motion along the radial and angular directions in the plane of the orbit occurs with the same periodicity, which leads to  conservation of the Runge-Lenz vector 
\begin{equation}
{\bf A} = {\bf p} \times {\bf L} - mQ \bf{ r/|r|},
\label{RL-vector}
\end{equation}
where ${\bf L}$ is the angular momentum. ${\bf A}$
points along the direction of the average electric dipole of the Kepler orbit.

\begin{figure}[t!]
\centering \includegraphics[width=0.85\columnwidth]{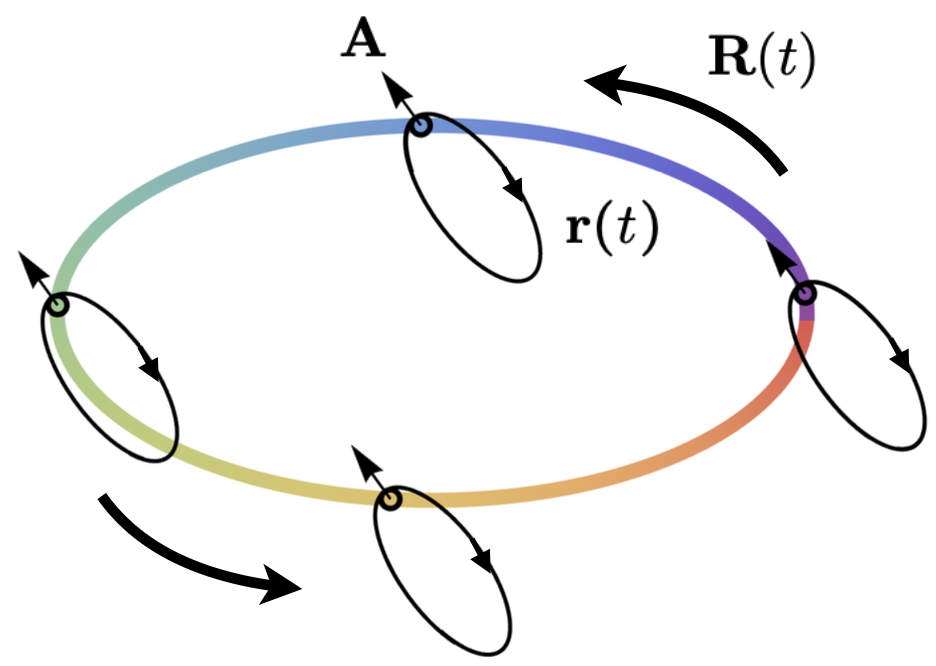}
\caption{A cyclic trajectory imposed for the vector ${\bf R}(t)$  (rainbow color; the time evolution is from blue to red), describing a slowly changing position of the nucleus (small circles). Small ellipses are the snapshots of numerically calculated Kepler orbits $\mathbf{r} (t)$  found at several different positions of the nucleus. The Runge-Lenz vector {\bf A} points from the nucleus towards the perigee of the orbit. For the Hamiltonian~(\ref{keplerH1}), the orientation of  ${\bf A}$ is an adiabatic invariant.}
\label{fig-Kepler}
\end{figure}

However, the high degree of symmetry in the standard Kepler problem also prevents the orbit from recording information about slow rotations of the central potential. Consider, for example, a three-dimensional Hamiltonian in which the position of the atomic nucleus, ${\bf R}(t)$, changes slowly with time:
\begin{equation}
H=\frac{{\bf p}^2}{2m} -\frac{Q}{|{\bf r}-{\bf R}(t)|}.
    \label{keplerH1}
\end{equation}
The numerically generated Fig.~\ref{fig-Kepler} shows that this parameter variation has no effect on the direction of the major axis of the elliptic electronic trajectory. In this sense, the Kepler orbit does not retain any memory of the nuclear rotational motion.

Instead, we search for conditions, under which similar cyclic changes of the parameters {\it can} change the direction of the electric dipole moment.
We begin with noting that a similar issue arises in classical 2-dimensional harmonic oscillator too: A slow translation of the parabolic potential minimum would not change individual adiabatic invariants (\ref{actions1}). 

Thus, we  recall the second condition required to
 induce a nontrivial geometric phase in the Foucault pendulum. Namely, this pendulum is a 3-dimensional system, in which the motion of the oscillator out-of-plane is suppressed, while the anisotropy axis is rotating. Precise strength of this anisotropy does not matter for the final geometric phase, as long as the rotation of the anisotropy axis is quasi-adiabatic. For example, the Foucault rotation angle is found in a 3-dimensional harmonic oscillator with an arbitrary strength of the uniaxial frequency anisotropy  \cite{Suzuki2025Geometric}. The mismatch between the in-plane and out-of-plane oscillation frequencies ensures that there is no transfer of energy from the in-plane to the out-of-plane direction during adiabatically slow rotations of the setup. Therefore, to induce a similar geometric phase in a Kepler orbit, we must introduce a  uniaxial anisotropy in addition to the Coulomb potential.
 

{\it Anisotropic Kepler problem}. Optical fields, used to localize Rydberg atoms, are also known  for distorting the spectrum of electronic Rydberg states in the range of $1-10$~MHz \cite{Topcu2013,Anderson2011}. Consider for example a fast AC-field along the $z$-axis induced by a standing wave of a laser beam:
\begin{equation}
{\bf E}(t)=\hat{z}E_0(z)\cos (\omega t).
    \label{AC-field}
\end{equation}
This force leads to the ponderomotive potential. For an electron with large principal number, a classical mechanical estimate for the strength of this potential is given by \cite{Knuffman2009}: 
\begin{equation}
U_{p}(z) = \frac{e^2}{4m \omega^2} E_0(z)^2.
 \label{Upond}   
\end{equation}
For a standing wave of the optical field, $E_0(z) = E\sin(kz)$, the potential minimum occurs at $z = 0$. Near this point, for small deviations of the orbit from the $xy$-plane, we find: 
$$
U_{p}(z)\approx \frac{\omega_0^2 m z^2}{2}, \quad \omega_0=\frac{ekE}{\sqrt{2}m\omega}.
$$

Thus, in the $xy$-plane the electron does not experience the positive shift of the potential energy, so we will call the $xy$ an easy plane. Within this plane, electronic motion is still described by the Kepler Hamiltonian. Due to the low electronic mass, estimates of $\omega_0$ for accessible optical fields indeed yield $\omega_0 \sim 10$~MHz, which agrees with experimental measurements of the corresponding Stark shift \cite{PhysRevLett.104.173001}. In what follows, we will assume that the time-dependent rotation of the electric field direction, about which we aim to leave a record in the Rydberg state, is much slower than the other relevant time scales. Hence, the time $1/\omega_0 \sim 0.1\ \mu$s sets the lower bound for the duration of the rotational motion we intend to detect.

This limit can be pushed closer to the inverse of the Kepler orbital period, $T_{\rm orb} \sim 1$~ns, either by employing much stronger beam intensities or by using infrared (smaller $\omega$ in Eq.~(\ref{Upond})) \cite{PhysRevA.49.R649,maclennan2021} rather than standard optical beam potentials. While this is possible, it would be more challenging to implement. Therefore, rotational dynamics occurring within the time interval of $1\ \mu$s to $1$~ms are the most accessible for experimental detection. This range already overlaps with frequencies of interest for future gyroscope applications.

The electronic Hamiltonian, including the ponderomotive potential, for an electron orbiting in the vicinity of the easy $xy$-plane of the Rydberg atom confinement has the form:
\begin{equation}
H=\frac{{\bf p}^2}{2m} - \frac{Q}{|{\bf r}|} +\frac{m \omega_0^2 z^2}{2}.
    \label{Ha2}
\end{equation}
We will consider initial conditions such that the elliptic orbits of all atoms are polarized in the easy plane along the $x$-axis. Experimentally, such conditions have been achieved previously \cite{Shao1997, EllipticStates1991}. However, anticipating a small misalignment, we should investigate whether it could lead to an unintended rotation of the main axis of the orbit.

\begin{figure}[t!]
\centering \includegraphics[width=1\columnwidth]{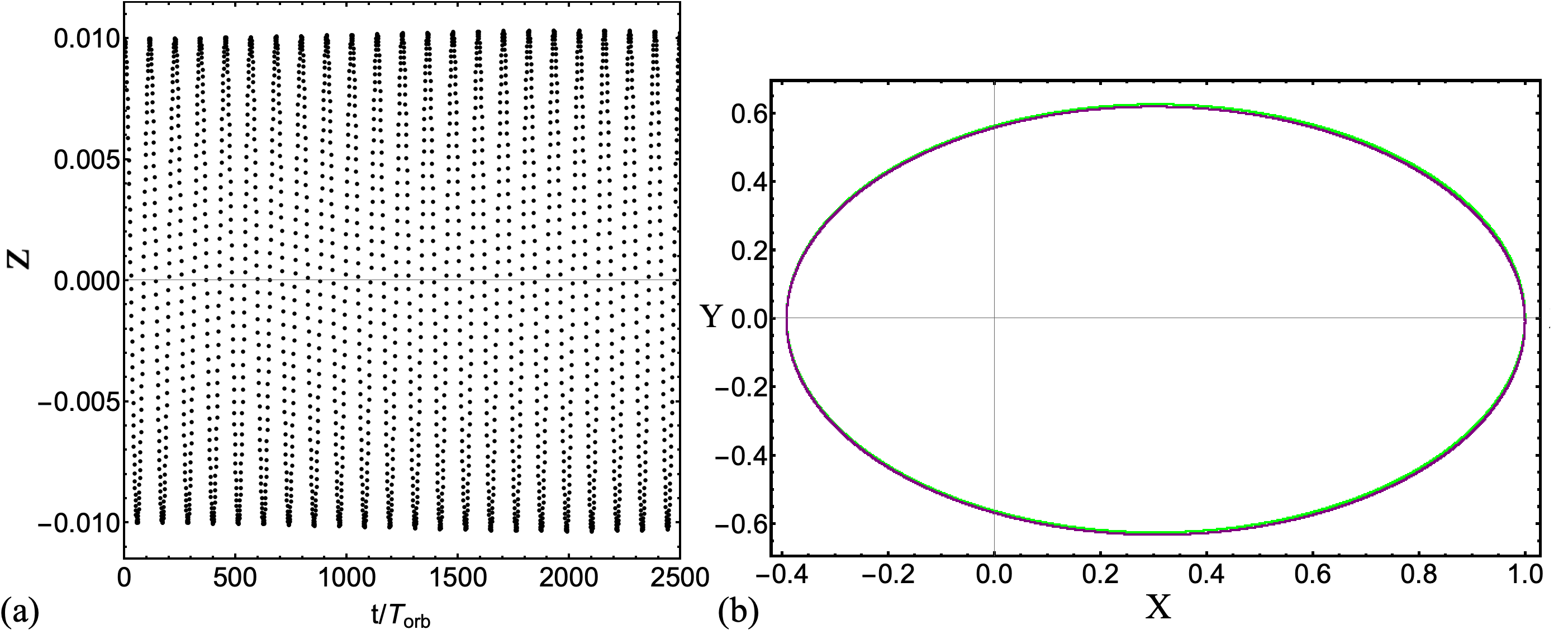}
\caption{(a) Evolution, according to the Hamiltonian~(\ref{Ha2}), of a small out-of-plane deviation of the electron trajectory from the easy plane. The trajectory in 3D-space was obtained by numerical simulations using Yoshida simplectic 3-rd order algorithm \cite{Yoshida1990, note}  for $m=Q=1$ and $\omega_0=0.2$, and then only the $z$-component was plotted. The motion started at ${\bf r}=(1.,0,0.01)$, ${\bf p}=(0,0.75,0)$, which corresponds to the time of one orbit cycle $T_{\rm orb}=3.65$. Each plotted point was obtained after an integer number of the orbiting periods: $n=t/T_{\rm orb}$, $n=0,1,\ldots$. (b) Projection of the  same trajectory on the easy $xy$-plane at $n=0$ (green) and after $n=3000$ cycles (purple). Within the simulation accuracy, there is no visible change of the direction of the main ellipse axis despite the time evolution is much larger than the period of oscillations along the $z$-axis and the deviation from the easy plane on the order of $|z|/|{\bf r}| \sim 1\%$.}
\label{z-fig}
\end{figure}

In Fig.~\ref{z-fig}, we show the evolution of the $z$-component of the electronic coordinate, transverse to the easy $xy$-plane. The trajectory was calculated for the Hamiltonian~(\ref{Ha2}), starting with an electron position and velocity almost but not precisely in the easy-plane, i.e., assuming that initially $|z| /|{\bf r}| \approx 0.01$. Figure~\ref{z-fig}(a) shows that, at constant values of the parameters, the last term in Eq.~(\ref{Ha2}) leads merely to relatively slow small amplitude oscillations of the $z$-component of ${\bf r}$ with zero mean value. In Fig.~\ref{z-fig}(b), we plot the projection of the elliptic trajectory on the easy plane right after the start (green) and after completion of 3000 Kepler orbit cycles (purple). This figure demonstrates that despite a small oscillatory deviation of the orbit from the easy plane, the electron dynamics is described by two degenerate frequencies, for the in-plane radial and angular motions, and a different  frequency, $\omega_0$, describing oscillations of the orbit along the $z$-axis.

The Hamiltonian~(\ref{Ha2}) is not globally super-integrable, but for the orbits in the easy plane, the radial and 
the angular oscillation frequencies are strictly degenerate. Figure~\ref{z-fig} demonstrates that, in addition, there is a range of misalignment angles of the orbits from the easy plane, for which the direction of the electric dipole is preserved during the time of experiment, so the effective Hamiltonian for such orbits is superintegrable up to corrections of a high power in $z/|r|$. These corrections are suppressed both by small values of $z/|r|$ and by fast oscillations of $z$ coordinate near its zero value, so they do not produce any visible effect on the dynamics within the experimentally relevant timescale. 

Thus, for analytical estimates of the geometric angle induced by the rotating anisotropy axis, the motion in the vicinity of the easy plane can be described by an effective 2-dimensional Kepler potential in the plane of the orbit, combined with a locally harmonic oscillator potential that induces relatively slow oscillations of the orbit's deviations from the easy plane:
\begin{equation}
H_{\rm eff}=\frac{{\bf p}_{\perp}^2}{2m}  -\frac{Q
}{|{\bf r}_{\perp}|} +\frac{P_z^2}{2m}+\frac{m \omega_0 Z^2}{2},
    \label{Ha3}
\end{equation}
where, we denoted ${\bf r}_{\perp} = (X, Y)$ and ${\bf p}_{\perp} = (P_x, P_y)$ as the coordinates and momenta of the electron in the instantaneous plane of the Kepler orbit, and $\omega_0$ as the frequency of out-of-plane oscillations induced by the ponderomotive force transverse to this plane. 

\begin{figure*}[t!]
\centering \includegraphics[width=2\columnwidth]{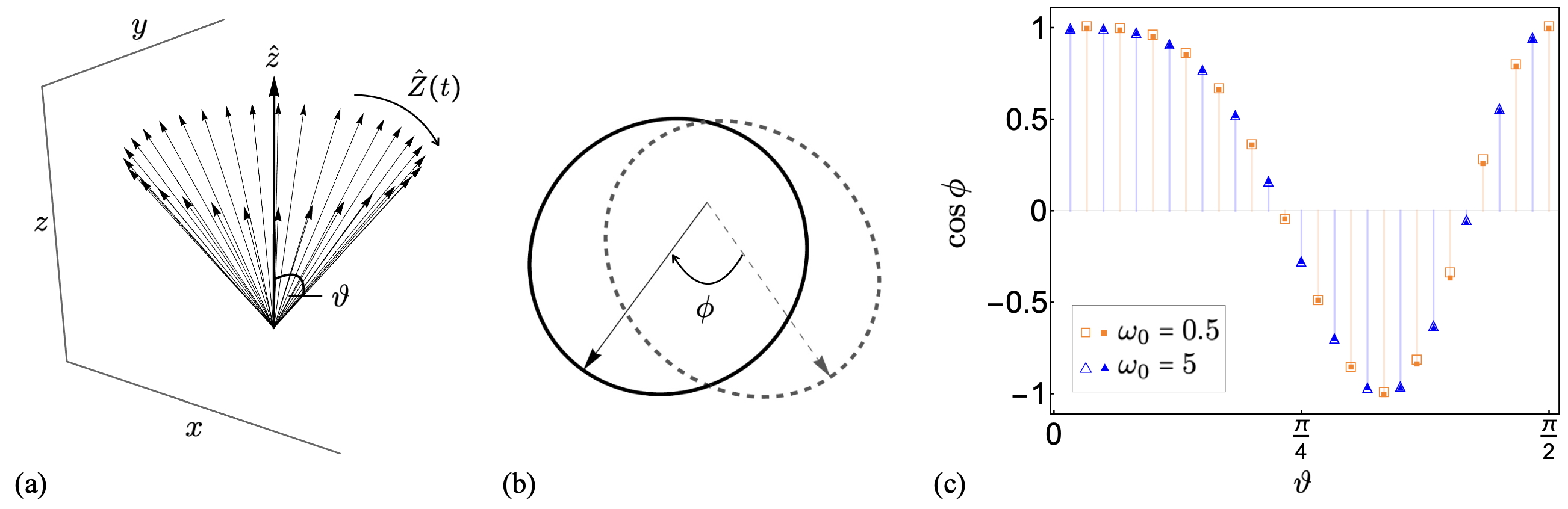}
\caption{(a) The adiabatic evolution of the field anisotropy vector $\hat{Z}(t)$, which subtends a solid angle $\Omega_{sa}$ given by Eq.~(\ref{sa}). The time-dependent Hamiltonian for simulations corresponds to Eq.~(\ref{Ha2}) with rotating anisotropy axis. It is written explicitly in \cite{note}. (b) The elliptic orbit at the start of the evolution (dashed curve) and at the end (solid curve). The angle $\phi$ between the main axes is predicted to coincide with  $\Omega_{sa}$. (c) This prediction (empty triangles/boxes), $\cos \phi = \cos(2\pi(1-\cos \vartheta))$. The blue triangles are the numerical results for simulations of the Hamiltonian equations of motion at strong anisotropy $\omega_0=5$. The orange squares are for the weak anisotropy, $\omega_0=0.5$. The protocol for a single point corresponds to a fixed $\vartheta$ in Eq.~(\ref{hatr}), and the angle $\varphi$  changing according to $\varphi=\pi [1+\tanh( t/\tau)]$, with $\tau\omega_0 \approx 10$ for $\omega_0=0.5$ and $\tau \omega_0 \approx 50$ for $\omega_0=5.$;  $t\in (-5\tau,5\tau)$. Other parameters and initial conditions are as in Fig.~\ref{z-fig}. }
\label{fig3}
\end{figure*}

{\it Rotating anisotropy axis}. 
 Let $\hat{x}$, $\hat{y}$, and $\hat{z}$ represent unit vectors along the axes of a fixed coordinate system and let the anisotropy axis rotate. Then, the unit vector $\hat{Z}$ along the direction of the AC-field can be parametrized by two time-dependent parameters, $\vartheta(t)$ and $\varphi(t)$: 
\begin{equation}
\hat{ Z}=\left(\begin{array}{c}
\sin \vartheta \sin\varphi \\
\sin \vartheta \cos \varphi\\
\cos \vartheta
\end{array}
\right).
    \label{hatr}
\end{equation}
Our goal is to compare the parameters of the electron's Kepler orbit before and after the anisotropy field vector $\hat{Z}$ completes a full cycle of its evolution.

It will be sufficient to consider the time-dependent protocol with  $\vartheta={\rm const}$, and $\varphi$ varying in the interval $\varphi \in [0,2\pi]$, as illustrated in Fig.~\ref{fig3}(a), so that the vector $\hat{Z}$ sweeps out a solid angle on the unit sphere equal to 
 \begin{equation}
 \Omega_{sa}=2\pi(1- \cos \vartheta).
 \label{sa}
 \end{equation} 
If the vector ${\hat{Z}}$ were not changing with time, the orbit would remain in the original easy plane. However, for slowly time-dependent parameters, the orbit adjusts to remain in the rotating easy plane, up to small out-of-plane nonadiabatic corrections, with  
$
|Z|\ll \sqrt{X^2+Y^2}
$,
which vanish after the completion of the driving protocol.
Thus, we find a situation very similar to that of the Foucault pendulum: The electron's motion is essentially governed by a super-integrable 2-dimensional Kepler Hamiltonian, but the plane of the orbit undergoes a slow rotation in 3-dimensional space.

{\it Cylindrical Coordinates}. In the plane of the orbit, we switch to cylindrical coordinates:
\begin{eqnarray}
\nonumber X &=& r \cos \phi, \quad Y = r \sin \phi, \\
p_r &=& \frac{X P_x + Y P_y}{r}, \quad p_{\phi} = X P_y - Y P_x.
\end{eqnarray}
The Hamiltonian in the rotating frame becomes
\begin{equation}
H_{\rm eff} = \frac{p_r^2}{2m} + \frac{p_{\phi}^2}{2 m r^2} - \frac{Q}{r}+\frac{m \omega_0 Z^2}{2} - p_{\phi} \dot{\varphi} \cos \vartheta,
\end{equation}
where the last term arises from the Coriolis force \cite{note}. We then switch to the action-angle variables \cite{landau1976mechanics}:
$$
I_{\phi}=p_{\phi}, \quad I_r=\frac{1}{2\pi} \oint p_r \, dr, 
$$
in which the Hamiltonian  is given by
\begin{equation}
H_{\rm eff}=-\frac{mQ^2}{2(I_r+I_{\phi})^2}+\frac{m \omega_0 Z^2}{2} -I_{\phi}\dot{\varphi}\cos \vartheta. 
\label{heff-fin}
\end{equation}
The first term in Eq.~(\ref{heff-fin}) is the standard form of the 2D Kepler Hamiltonian written in action-angle variables \cite{landau1976mechanics}. The effect of  the second term on the orbit is merely to keep this orbit within the easy plane with $Z=0$, as for the plane of oscillations of the Foucault pendulum.
This Hamiltonian does not depend on the angle variables canonically conjugate to $I_{\phi}$ and $I_r$, which we denote as $\theta_{\phi}$ and $\theta_r$, respectively. The difference between these phases, $\delta \theta = \theta_r - \theta_{\phi}$, changes with time according to 
$$
\frac{d\, \delta \theta }{dt} =\frac{\partial H}{\partial I_r}-\frac{\partial H}{\partial I_{\phi}} = \cos \vartheta \frac{d \varphi}{dt}. 
$$
Integrating over the evolution along a closed path traced by $\hat{Z}(t)$, we obtain the shift, known as the Hannay angle \cite{Hannay1985}, of the phase between radial and angular oscillations after completion of the anisotropy axis rotation:
\begin{equation}
\delta \theta = \oint_{\varphi(t)} \cos \vartheta \, d\varphi = 2\pi \cos \vartheta,
\label{delta }
\end{equation}
where the integral is taken over a closed path on a unit sphere parametrized by the angles $\vartheta$ and $\varphi(t)$.

Let us count the angle variable $\theta_r$ starting from the apogee of the orbit, and identify $\theta_{\phi}$ with the angle $\phi$ in cylindrical coordinates. Then, the orbit with $\delta \theta = 0$ corresponds to the apogee point on the $x$-axis of the initial orbit. For an orbit with a finite $\delta \theta$, we have
$\theta_r = \theta_{\phi} + \delta \theta$. The value $\theta_r = 0$ then corresponds to an angle in cylindrical coordinates given by:
\begin{equation}
\phi = \theta_{\phi}=-\delta \theta.
\label{rota1}
\end{equation}
Thus, we find that upon completion of the rotation of the anisotropy axis, the average dipole moment of the Rydberg atom ends up being rotated by an angle given by Eq.~(\ref{rota1}), as shown in Fig.~\ref{fig3}(b).

In the rotating frame, according to Eq.~(\ref{rota1}), the rotation angle of the main axis of the orbit is $\phi = -2\pi \cos \vartheta$. Since the rotating frame makes a full rotation around the fixed $\hat{z}$-axis by an angle of $2\pi$, in the fixed frame the electric dipole moment rotation angle is given by
\begin{equation}
\phi=\Omega_{sa}.
\label{phi-fin}
\end{equation}
The geometric nature of the result (\ref{phi-fin}) also means that it is valid for an arbitrary time-dependent closed control protocol $(\vartheta(t), \varphi(t))$, because any solid angle subtended by such a path can be covered by infinitesimal contours for which we have already proven the validity of Eq.~(\ref{phi-fin}).

Figure~\ref{fig3} shows the results of our numerical test of Eq.~(\ref{phi-fin}), for different strengths of the anisotropy frequency $\omega_0$ and the colatitude angle $\vartheta$ \cite{note}. The angle $\varphi$ was varied according to  $\varphi=\pi [1+\tanh( t/\tau)]$, where $\tau \gg 1/\omega_0$, which guaranteed the applicability of the adiabatic approximation. 
The cosine of the angle $\phi$ was found as a scalar product between the initial and final directions of the unit vector pointing from the center to the apogee. The results were in excellent agreement with Eq.~(\ref{phi-fin}) as long as  $\tau$ was at least an order of magnitude longer than the characteristic timescale $1/\omega_0$ of the out-of-plane oscillations shown in Fig.~\ref{z-fig}(a).

{\it Discussion.} In a laboratory, this effect can be demonstrated by placing Rydberg atoms initially at the minimum of the ponderomotive potential, as in \cite{Zhang2011Magic,Moore2015Ponderomotive}, and then slowly modulating the direction of the optical beam periodically. Alternatively, one can create a static, spatially spiraling optical field using chiral optics \cite{Uesugi2022}. Rydberg atoms flying along the potential minimum of this spiral \cite{RevModPhys.94.041001,Traxler2012} would then experience a rotating uniaxial anisotropy.

We anticipate potential applications, in which the rotation of the anisotropy axis is achieved when the entire setup is rigidly attached to the walls of a moving structure, such as a rocket experiencing sub-millisecond perturbations and torques. Even though the AC-field has a fixed direction in the setup's frame, mechanical disturbances induce rotations of the frame itself. In contrast to Fig.~\ref{fig-Kepler}, these rotations result in a residual rotation of the electric dipole moment, which can accumulate with time. After the frame's orientation is restored, the new dipole direction retains the memory of the prior rotation, and thus can be used to adjust the motion of the rocket.

Geometric phases have been rarely addressed in the literature on Rydberg states. Notable examples include a geometric shift at a wavepacket revival, predicted in chaotic semiclassical dynamics \cite{Jarzynski1995GeoPhase}, and an approach to generating quantum gates for qubits using the Berry phase \cite{wu2017rydberg}. A key distinction of the geometric angle discussed here is that 
it does not arise as a small nonadiabatic correction to a rapidly varying dynamical contribution to an angular variable. In fact, the dipole moment direction does not change under static parameters. Therefore, the geometric angle  is the only effect responsible for altering the dipole moment direction in our setup. 
In this sense, the angle shares many characteristics with non-Abelian quantum mechanical geometric phases, which arise within an energy-degenerate subspace \cite{wilczek1984nonabelian}.

This geometric angle is also independent of the electronic orbital angular momentum and the speed of the frame's rotation, provided that the adiabaticity conditions are satisfied. It is robust against variations in the anisotropic field frequency $\omega_0$ and other modifications to the anisotropy potential -- this potential is needed only to create an easy plane with an effectively two-dimensional Kepler Hamiltonian. 
This phase  does not depend on the initial conditions for the quickly changing angle variables, so there is no need to create a localized wave packet in order to observe it.  Since the effect does not directly rely on quantum interference or other forms of quantum many-body correlations, it should also be robust against moderate losses of the Rydberg states when experiments are performed with many atoms simultaneously.

A possible approach to finally measure this angle is to make atoms flying through a nonuniform electric field, which deflects Rydberg atoms by an angle that depends on the relative directions of the field gradient and the electric dipole polarization, as in experiments \cite{PhysRevA.76.023405,Hogan2016EPJTI}. Another approach is to measure a DC Stark shift induced by an additional electric field applied along the original electric dipole polarization in the absence of the ponderomotive potential.  The energy of interaction of an electric dipole moment with an electric field depends on the cosine  of the angle between them, which is reflected in the shift of an absorption frequency peak \cite{Zhelyazkova2015Dipole}.   

\begin{acknowledgements}
\noindent
This work was supported primarily by the U.S. Department of Energy, Office of Science, Office of Advanced Scientific Computing Research, through the Quantum Internet to Accelerate Scientific Discovery Program, F.S. acknowledges support from the Los Alamos National Laboratory LDRD program under project number 20230049DR and the Center for Nonlinear Studies under project number 20220546CR-NLS.
\end{acknowledgements}
\
\bibliographystyle{apsrev4-2}
\bibliography{ref}

\begin{appendix}

\section{Yoshida's algorithm}
\label{appA}

In Hamiltonian mechanics, preserving the symplectic structure of phase space is crucial for accurate long-term simulations. Symplectic integrators ensure that energy and other invariants are conserved over long time spans better than non-symplectic methods. One popular second-order symplectic integrator is the \textit{Velocity Verlet} method. To achieve higher-order accuracy while retaining symplectic properties we used Yoshida's method.

Yoshida's algorithm \cite{Yoshida1990}  constructs higher-order symplectic integrators by composing second-order integrators with specially chosen coefficients. For a Hamiltonian of the form
\[
H(p, q) = T(p) + V(q),
\]
the time evolution can be split into kinetic and potential updates.

Let $S_2(\Delta t)$ be a second-order symplectic integrator (Velocity Verlet in our case). Then, a third-order symplectic integrator is defined as a composition:

\[
S_3(\Delta t) = S_2(a_1 \Delta t) \circ S_2(a_2 \Delta t) \circ S_2(a_1 \Delta t),
\]
where the coefficients $a_1$ and $a_2$ are:
\[
a_1 = \frac{1}{2 - 2^{1/3}}, \quad
a_2 = -\frac{2^{1/3}}{2 - 2^{1/3}}.
\]

These coefficients are chosen to cancel out error terms up to $\mathcal{O}(\Delta t^4)$ while preserving the symplectic nature of the integrator.

\section{Kepler Hamiltonian in a rotating frame}
\label{appB}

Let us find the transformation from the fixed to the rotated frame for the Hamiltonian $H_{\rm eff}$ in Eq.~(7) in the main text.  
Due to the rotational symmetry of interactions around $\hat{Z}$, the axes in plane transverse to $\hat{Z}$ can be chosen arbitrarily, with the only requirements that corresponding unit vectors along them, $\hat{ \varphi}$ and $\hat{ \vartheta}$, must be mutually orthogonal with each other and $\hat{r}$, and varying periodically with $\varphi$. Thus, we choose them to be
\begin{equation}
\hat{\varphi}=\left(\begin{array}{c}
 \cos \varphi \\
- \sin \varphi\\
0
\end{array}
\right), \quad \hat{ \vartheta}=\left(\begin{array}{c}
\cos \vartheta \sin \varphi \\
\cos \vartheta \cos \varphi\\
-\sin \vartheta
\end{array}
\right).
    \label{hatangles}
\end{equation}

In the frame with the axes along these vectors, the joint coordinates and momenta are given by 
\begin{eqnarray}
\nonumber{\bf r} &=&X\hat{\varphi}+Y\hat{\vartheta}+Z\hat{Z},\\ 
\label{rp-vec}
{\bf P} &=&P_x\hat{\varphi}+P_y\hat{\vartheta}+P_z\hat{Z},
\end{eqnarray}

Let, in the fixed frame, the same coordinate and momentum vectors have components, respectively, ${\bf r}_0=(x,y,z)$ and ${\bf p}_0=(p_x,p_y,p_z)$. From Eqs.~(8), (\ref{hatangles}), and (\ref{rp-vec}), we find then the  transformation that separates the variables of the rotated structure Hamiltonian:
\begin{equation}
\left(\begin{array}{c}
x\\
y\\
z
\end{array}
\right)=R(\vartheta,\varphi) \left(\begin{array}{c}
X\\
Y\\
Z
\end{array}
\right), \quad 
\left(\begin{array}{c}
p_x\\
p_y\\
p_z
\end{array}
\right)=R(\vartheta,\varphi) \left(\begin{array}{c}
P_x\\
P_y\\
P_z
\end{array}
\right),
\label{rot-tr}
\end{equation}
where $R(\vartheta, \varphi)$ is a rotation matrix given explicitly by
\begin{equation}
R(\vartheta, \varphi) =
\begin{bmatrix}
\cos \varphi & \cos \vartheta \sin \varphi & \sin \vartheta \sin \varphi \\
- \sin \varphi & \cos \vartheta \cos \varphi & \sin \vartheta \cos \varphi \\
0 & - \sin \vartheta & \cos \vartheta.
\end{bmatrix}
\label{Rmt}
\end{equation}

When the angle $\varphi$ slowly varies with time, the Hamiltonian acquires a nonadiabatic correction
$$
{\cal H}= H({\bf r}_0({\bf r},{\bf P}),{\bf p}_0({\bf r},{\bf P}))+\delta H ({\bf r},{\bf P}),
$$
where $H({\bf r}_0({\bf r},{\bf P}),{\bf p}_0({\bf r},{\bf P}))$ has the form (7)
$$
 H({\bf r}_0({\bf r},{\bf P})) = \frac{P_x^2+P_y^2+P_z^2}{2m} - \frac{Q}{\sqrt{X^2+Y^2}} + \frac{\omega_0^2m Z^2}{2}, 
$$
and  $\delta H$ is found from the invariance of the action functional: 
$$
S=\int \{{\bf p}_0 \, d{\bf r}_0-H({\bf r}_0,{\bf p}_0)\,dt \},
$$
leading to
\begin{eqnarray}
\nonumber \delta H ({\bf r},{\bf P})&=&-{\bf p}_0({\bf r},{\bf P})\cdot\partial_t{\bf r}_0({\bf r},{\bf P}) = -{\bf P} \cdot R^{T} \frac{dR}{d\varphi} {\bf r} \dot{\varphi}\\
\label{Fhprime}
\nonumber &=&\{\cos \vartheta (P_yX-P_xY)-\sin \vartheta (P_xZ-P_zY)\}\dot{\varphi}.
\end{eqnarray}
The term $\propto \sin \vartheta$ in $\delta H$ mixes variables that oscillate with different frequencies. Therefore, in the adiabatic limit, it can be disregarded, as discussed in standard textbooks \cite{landau1976mechanics}. Switching to the action-angle variables, the remaining correction, $\propto \cos \vartheta$, leads to the Coriolis force contribution for the rotated frame.

\section{Hamiltonian dynamics simulated for Fig.~3}
\label{appC}

In the frame whose $Z$-axis coincides with the direction of anisotropy, the Hamiltonian has the form 
\begin{equation}
   H= \frac{{\bf p}^2}{2} - \frac{1}{|{\bf r}|} + \frac{\omega_0^2 Z^2}{2}, 
   \label{hsim}
\end{equation}
where ${\bf r}=(X,Y,Z)$,  and the Kepler potential is in 3-dimensional space. 
In the fixed coordinates $(xyz)$, the out-of-plane coordinate in the rotated plane is give by
$$
Z=z\cos\vartheta +y\sin \vartheta \cos \varphi +x \sin \vartheta \sin \varphi
$$

The first two terms in (\ref{hsim}) are symmetric under the frame rotation, so we  simulated the canonical equations of motion in the fixed frame with an explicitly time-dependent Hamiltonian 
\begin{equation}
   H= \frac{{\bf p}^2}{2} - \frac{1}{|{\bf r}|} + \frac{\omega_0^2 (z\cos\vartheta +y\sin \vartheta \cos \varphi +x \sin \vartheta \sin \varphi)^2}{2}, 
   \label{hsim2}
\end{equation}
where both $\vartheta$ and $\varphi$ could be slowly time-dependent. 

We started at initial conditions for which the electron's coordinate and velocity were in-plane transverse to $\hat{Z}$-axis. Specifically, that initially

$$
(x,y,z)=(R_0,0,0), \quad (v_x,v_y,v_z)=V_0(0,\cos \vartheta, -\sin \vartheta)
$$
for certain $R_0$ and $V_0$ that set parameters of the orbit.

\end{appendix}
\end{document}